\documentclass[reprint,prl]{revtex4-2}
\usepackage{graphicx}
\usepackage{microtype}              
\usepackage[utf8]{inputenc}         
\usepackage[T1]{fontenc}            
\usepackage{newtxtext}              
\usepackage[smallerops]{newtxmath}  
\usepackage[dvipsnames]{xcolor}
\usepackage{hyperref}
\hypersetup{
    colorlinks=true,
    linkcolor=NavyBlue,
    citecolor=NavyBlue,
    urlcolor=NavyBlue
}
\def\iid{\emph{iid}}
\def\secret{\sigma^m}
\def\shares{\sigma^s}
\def\share{\tau}
\def\hidden{\sigma^h}
\def\visible{\sigma^v}
\begin{document}\frenchspacing
\title{Random-Energy Secret Sharing via Extreme Synergy}
\author{Vudtiwat Ngampruetikorn}
\author{David J.\ Schwab}
\affiliation{%
    Initiative for the Theoretical Sciences
    and
    CUNY--Princeton Center for the Physics of Biological Function,
    The Graduate Center, CUNY,
    New York, NY 10016
}
\begin{abstract}
The random-energy model (REM), a solvable spin-glass model, has impacted an incredibly diverse set of problems, from protein folding to combinatorial optimization to many-body localization. Here, we explore a new connection to secret sharing. We formulate a secret-sharing scheme, based on the REM, and analyze its information-theoretic properties. Our analyses reveal that the correlations between subsystems of the REM are highly synergistic and form the basis for secure secret-sharing schemes. We derive the ranges of temperatures and secret lengths over which the REM satisfies the requirement of secure secret sharing. We show further that a special point in the phase diagram exists at which the REM-based scheme is optimal in its information encoding. Our analytical results for the thermodynamic limit are in good qualitative agreement with numerical simulations of finite systems, for which the strict security requirement is replaced by a tradeoff between secrecy and recoverability. Our work offers a further example of information theory as a unifying concept, connecting problems in statistical physics to those in computation. 
\end{abstract}
\maketitle

Keeping sensitive information from bad actors has long been a challenge. Lock a secret in a safe with one master key and the secret is gone forever should the key be lost. Make copies of the key and the risk is of one of them falling into the wrong hands. Secret sharing offers a solution~\cite{shamir:79,blakley:79}. In a $(k,n)$ threshold scheme, a secret is split into $n$ shares. Reading the secret requires $k$ or more distinct shares, and $k\!-\!1$ or fewer shares reveal absolutely no information about the secret. This strategy allows secure information storage whose degrees of security and fail safety are tunable through $k$ and $n$. Moreover, the threshold requirement provides a mechanism for enforcing coordination and agreement between parties with different shares, thus making secret-sharing schemes an important building block in distributed computing, see, e.g., Ref~\cite{beimel:11}.

\begin{figure}
\centering
\includegraphics[width=\linewidth]{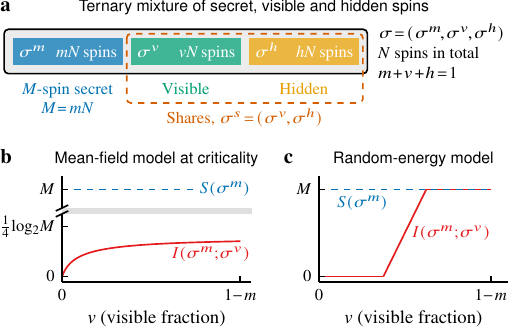}
\caption{\label{fig:synergy}%
The random-energy model exhibits extremely strong and highly synergistic correlations.
(a)~We divide $N$ spins into three disjoint groups, $\sigma\!=\!(\secret\!,\visible\!,\hidden)$. The secret $\secret$ consists of $M\!=\!mN$ spins. The other spins are the shares $\shares\!=\!(\visible\!,\hidden)$, of which $vN$ spins are visible $\visible$ and $hN$ are hidden $\hidden$\!. We consider the reconstruction of the secret $\secret$ from an observation of the visible spins $\visible$\!. 
(b\,\&\,c)~The amount of information in the secret is quantified by its entropy $S(\secret)\!\sim\!M$ bits (dashed). The information that the visible spins have about the secret is measured by their mutual information $I(\secret;\visible)$ (solid), here depicted as a function of the visible fraction $v$ at a fixed secret fraction $m$. 
(b)~For the fully connected Ising model at criticality, the information increases with more visible spins. This increase diminishes as $v$ grows, indicating \emph{redundant} coding of secret information. The logarithmic scaling of the information with $N$ signifies strong correlations associated with critical behaviors; away from the critical point, the information does not grow with $N$. 
(c)~For the random-energy model in the paramagnetic phase ($T\!=\!\sqrt{2}T_c$), the information becomes positive only with enough visible spins. Moreover, the information is \emph{extensive} in this case, indicating even stronger correlations than those in typical critical systems. Importantly, this extensivity means that visible spins can encode \emph{all} of the secret information, thus allowing perfect secret reconstruction. This behavior is a signature of extreme synergy among the spins---that is, while individual spins leak no secret information, an adequately large collective of spins can completely reveal the secret.%
}
\end{figure}

While constructions of secret sharing schemes are well known~\cite{shamir:79,blakley:79}, such schemes seem too contrived to appear in nature. A primary reason for this is that secret sharing requires a potentially extreme form of \emph{synergy}, where information is stored in the joint state of a collection of random variables that is absent from its subsets~(see Fig~\ref{fig:synergy}). Yet, some well-known models of physical and biological systems exhibit similar behaviors. For example, information in the interior of black holes is encoded in the entanglement between Hawking radiation subsystems; as a result, information recovery is possible only when all radiation parts are available~\cite{balasubramanian:21}. Another example is combinatorial coding in an ensemble of neurons, in which two neurons together can carry many more bits about a stimulus than the sum of their individual information~\cite{brenner:00,schneidman:03b}. In this case, the neural code represents a secret-sharing scheme, albeit an imperfect one since each individual neuron can still be predictive of the stimulus. Here, we show that a spin-glass model, namely the random-energy model~\cite{derrida:80,derrida:81}, also implements a secret-sharing scheme.

The random-energy model (REM) has proved remarkably versatile. It offers a minimal model and useful theoretical benchmarks for a wide range of problems, from protein folding~\cite{bryngelson:87,shakhnovich:90,pande:96} to self assembly~\cite{hormoz:11} to biodiversity~\cite{fisher:14} to many-body localization~\cite{laumann:14,lazarides:15,baldwin:16}. It also connects statistical physics with concepts in computation such as error correction~\cite{sourlas:89} and combinatorial optimization~\cite{mertens:00,borgs:01} (see Ref~\cite{mezard:09} and references therein).

In this Letter, we investigate yet another intriguing connection, between the REM and cryptography. We analyze the information-theoretic properties of the REM and show that the correlations between its subsystems satisfy the information-theoretic requirement of secure secret-sharing schemes. We map the REM to $(k,n)$ threshold schemes and derive a phase diagram of secure regions. We show that for every $(k,n)$, a special set of model parameters exists such that the encoding of secret information in each of the $n$ shares is at the physical limit and thereby optimal. In addition, we numerically compute the relevant information terms for finite-spin REMs and demonstrate that they also implement secret-sharing schemes, albeit with the security requirement replaced by a tradeoff between secrecy and recoverability.

The random-energy model describes a system of $N$ Ising spins---$\sigma\!=\!(\sigma_1,\sigma_2,\dots,\sigma_N)$ with $\sigma_i\!=\!\pm1$---whose energy levels $E_\sigma$ are \iid\ Gaussian variables. That is, the probability of a system configuration $\sigma$ reads
\begin{equation}\label{eq:REM}
    P(\sigma) = e^{-\beta E_\sigma}/Z
    \quad\text{with}\quad 
    E_\sigma \sim \mathcal{N}(0,NJ^2/2),
\end{equation}
where $\beta$ denotes the inverse temperature and $J$ is an intensive parameter. In the limit $N\!\to\!\infty$, the partition function admits an analytical form~\cite{derrida:80,derrida:81}
\begin{equation}
    \ln Z = \ln \sum_{\sigma}e^{-\beta E_\sigma} 
    = N\ln2\times \left\{%
    \begin{array}{ll}
         1+(\frac{\beta}{\beta_c})^2    &\text{if $\beta\le\beta_c$}  \\
         2\frac{\beta}{\beta_c}         &\text{if $\beta>\beta_c$.}
    \end{array}%
    \right.
\end{equation}
The critical point $\beta_cJ\!=\!2\sqrt{\ln2}$ marks a first-order phase transition between a paramagnetic state at high temperatures $\beta\!<\!\beta_c$, and a \emph{frozen} state at low temperatures $\beta\!>\!\beta_c$. We can obtain all thermodynamic variables from the above partition function. Of particular interest is the entropy (in bits)~\footnote{We slightly abuse the notation and use $\sigma$ to denote both random variables and their realizations.}
\begin{equation}\label{eq:entropy}
    S(\sigma)
    = \log_2 Z - \beta\frac{\partial\log_2 Z}{\partial\beta}
    = N\times\left\{%
    \begin{array}{ll}
         1-t^{-2}   &\text{if $t\ge1$}  \\
         0          &\text{if $t<1$.}
    \end{array}%
    \right.
\end{equation}
Here and throughout, we let $t\!\equiv\!\beta_c/\beta$ denote the reduced temperature. In the high-temperature limit, thermal noise dominates and the system becomes uncorrelated and random, resulting in one bit of entropy per spin. As the temperature drops, the entropy decreases until it vanishes at the critical point below which only an $O(1)$ number of least energetic configurations prevail.

\begin{figure}
\centering
\includegraphics[width=\linewidth]{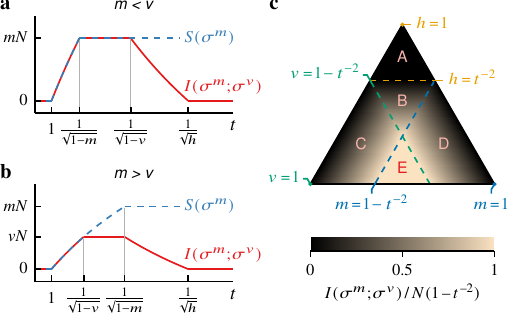}
\caption{\label{fig:ternary}%
Mutual information between subsystems \emph{vs} temperature and system composition. 
(a\,\&\,b)~The temperature dependence of the secret entropy (dashed) and the mutual information between the secret and visible spins (solid), see Fig~\ref{fig:synergy}a. The secret entropy $S(\secret)$ vanished in the frozen phase $t\!<\!1$ and grows with $t$ until it plateaus at $M$ bits [Eq~\eqref{eq:marginal_entropy}]. The information $I(\secret;\visible)$ exhibits similar behaviors at low temperatures, vanishing for $t\!<\!1$ and increasing with $t$ near the onset of the paramagnetic phase. But this information is bounded by either the entropy of the secret or that of visible spins whichever smaller; as a result, it saturates at $\min(m,v)N$ bits [Eq~\eqref{eq:info}]. At high temperatures, thermal noise dominates and the information decreases with $t$, approaching zero at $t\!=\!1/\sqrt{h}$. 
(c)~This information depends on the composition of the secret, visible and hidden spins, parametrized by their fractions $(m,v,h)$ (Fig~\ref{fig:synergy}a). For $t\!>\!1$, this ternary diagram has five regions, A-E. The information density, $I(\secret;\visible)/N$, is zero in A, $t^{-2}\!-\!h$ in B, $m$ in C, $v$ in  D and $1\!-\!t^{-2}$ in E. For $t\!>\!\sqrt{2}$, Region E disappears. See main text for details.%
}
\end{figure}

To turn the REM into a secret-sharing scheme, we divide the spins into two disjoint groups $\sigma\!=\!(\secret,\shares)$: the secret message $\secret$ consists of $M$ spins, and the other spins $\shares$ represent all of the shares that will be distributed to different parties, see Fig~\ref{fig:synergy}a. The information content of the secret is quantified by its entropy
\begin{equation}
    S(\secret) = S(\secret,\shares) - S(\shares\mid\secret),
\end{equation}
where $S(\secret,\shares)\!=\!S(\sigma)$ is the entropy of the entire system, given by Eq~\eqref{eq:entropy}, and $S(\shares\,|\,\secret)$ the conditional entropy. To compute the latter, we note that the energy levels are \iid\ random variables and thus fixing $\secret$ leaves the energy-level statistics unchanged. In other words, the conditional model is defined by the same energies as the original model, i.e., $P(\shares\,|\,\secret)\!\propto\!e^{-\beta E_{\shares|\secret}}$ with $E_{\shares|\secret}\!\sim\!\mathcal{N}(0,NJ^2/2)$. We see that the conditional system is also an REM, but with $\tilde N\!=\!N\!-\!M$ spins and a variance parameter $\tilde J\!=\!\sqrt{N/(N\!-\!M)}J$ (such that the energy variance is unchanged, i.e., $\tilde N \tilde{J}^2\!=\!NJ^2$). As a result, its critical point is given by $\tilde\beta_c\!=\!\sqrt{1\!-\!m}\beta_c$, and the conditional entropy by [see Eq~\eqref{eq:entropy}]
\begin{equation}\label{eq:cond_entropy}
    S(\shares\mid\secret)
    = N\times\left\{%
    \begin{array}{ll}
         1-m-t^{-2}     &\text{if $t\ge1/\sqrt{1-m}$}  \\
         0              &\text{if $t<1/\sqrt{1-m}$,}
    \end{array}%
    \right.
\end{equation}
where $m\!=\!M/N$. Combining Eqs~(\ref{eq:entropy}-\ref{eq:cond_entropy}) yields~\footnote{\label{ft:marginal_entropy}The argument, leading to this expression, extends to any extensive subsystem of the REM; in particular, Eq~\eqref{eq:marginal_entropy} gives the marginal entropy of a subsystem that is a fraction $m$ of the full system.}
\begin{equation}\label{eq:marginal_entropy}
    S(\secret) 
    = 
    \min[S(\sigma),M]
    =
    N\times\left\{%
    \begin{array}{ll}
        \min(m,1\!-t^{-2})  &\text{if $t\ge1$}  \\
        0                   &\text{if $t<1$.}
    \end{array}%
    \right.
\end{equation}
We see that the entropy of the secret, like that of the entire system, vanishes in the frozen phase $t\!<\!1$ and becomes finite in the paramagnetic phase $t\!>\!1$. Quite remarkably, this entropy is \emph{identical} to that of the entire system for $1\!\le\!t\!\le\!1/\sqrt{1\!-\!m}$; that is, different parts of the system are maximally correlated. But while the system's entropy approaches $N$ bits as $t\!\to\!\infty$ [Eq~\eqref{eq:entropy}], the secret entropy plateaus at $M$ bits for $t\!>\!1/\sqrt{1\!-\!m}$. Importantly, the information content of the secret reaches the maximum capacity of $M$ bits at a finite temperature $t\!=\!1/\sqrt{1\!-\!m}$ (see also Fig~\ref{fig:ternary}a-b).

We now turn to the recoverability of the secret. To this end, we further split the shares into two parts, $\shares\!=\!(\visible,\hidden)$. The visible spins $\visible$ amount to a fraction $v$ of the system and the hidden spins $\hidden$ to a fraction $h$. The secret, visible and hidden spins together make up the system, i.e., $\sigma\!=\!(\secret,\visible,\hidden)$ and $m+v+h\!=\!1$, see Fig~\ref{fig:synergy}a.

Suppose we only have access to the visible spins. Reconstructing the secret from these spins is equivalent to an inference problem, characterized by the mutual information,
\begin{equation}
    I(\secret;\visible)=S(\secret)-S(\secret\mid\visible),
\end{equation}
which measures the reduction in the uncertainty, quantified by entropy, of the secret once we have observed the state of the visible spins. Equation \eqref{eq:marginal_entropy} provides the expression for the secret entropy $S(\secret)$. To obtain the conditional entropy, we recall $S(\secret\,|\,\visible)\!=\!S(\secret,\visible)\!-\!S(\visible)$. Since $\visible$ is a fraction $v$ of the system, its marginal entropy is given by Eq~\eqref{eq:marginal_entropy} but with $v$ in place of $m$, i.e., $S(\visible)\!=\!\min[S(\sigma),vN]$. And similarly $S(\secret,\visible)\!=\!\min[S(\sigma),(m\!+\!v)N]$. As a result, we have~\footnote{The mutual information between two equal halves of the REM \cite{bialek:20} is a special case of our result, with $h\!=\!0$ and $m\!=\!v\!=\!1/2$.}
\begin{equation}\label{eq:info}
    \!\!
    I(\secret;\visible)
    =N\times\left\{%
    \begin{array}{lr}
        \multicolumn{2}{l}{[\min(m,1-t^{-2})+\min(v,1-t^{-2})}\;\;\;\\
        \;\;-\min(m+v,1-t^{-2})]    &\text{if $t\ge1$\phantom{.}}  \\
        0                           &\text{if $t<1$.}
    \end{array}%
    \right.\!
\end{equation}
We note that this mutual information is \emph{extensive}. This behavior is in contrast with typical statistical physics models, in which the information between macroscopic parts of the system grows with the system size only at critical points and only subextensivel~\cite{grassberger:86,bialek:01,wilms:12,cohen:15,ngampruetikorn:23} (see also Fig~\ref{fig:synergy}b-c). The extensivity of this information signifies unusually strong correlations and implies that the REM is \emph{unlearnable} from finite measurements~\cite{bialek:20}.

We depict the temperature dependence of this information in Fig~\ref{fig:ternary}a-b. Like entropy, it increases from zero as $\sim\!1\!-\!t^{-2}$ as the system leaves the frozen phase and enters the paramagnetic phase. But unlike entropy, it decreases at high temperatures as $\sim\!t^{-2}\!-\!h$ until vanishing at $t\!=\!1/\sqrt{h}$ (with $h\!=\!1\!-\!m\!-\!v$). At intermediate temperatures, it plateaus at either $mN$ or $vN$ bits, whichever is less. This plateau results from the upper bound of mutual information $I(A;B)\!\le\!\min[S(A),S(B)]$ for discrete random variables $A$ and $B$. In addition, we see that for $t\!\le\!1/\sqrt{1\!-\!v}$ the visible spins encode \emph{all} of the information the secret spins carry, i.e., $I(\secret;\visible)\!=\!S(\secret)$. In particular, when $m\!<\!v$, the information plateau coincides with that of the secret entropy (Fig~\ref{fig:ternary}a); in other words, the secret reaches its maximum information capacity (of $M$ bits) and all of that is encoded in the visible spins.

Figure~\ref{fig:ternary}c illustrates how the mutual information between secret and visible spins depends on the fractions $m$, $v$ and $h$ in the paramagnetic state ($t\!>\!1$). This information vanishes in Region A where a large part of the system is hidden $h\!>\!t^{-2}$. This region corresponds to the high-temperature limit $t\!>\!1/\sqrt{h}$ in Fig~\ref{fig:ternary}a-b. Decreasing the hidden fraction pushes the ternary mixture into Region B, where the information becomes finite at $(t^{-2}\!-\!h)N$ bits. This region is the only one that grows with temperature. The others shrink; in particular, Region E completely disappears for $t\!>\!\sqrt2$. When the secret fraction dominates $m\!>\!1\!-\!t^{-2}$, the information is equal to the entropy of visible spins, $vN$ and $(1\!-\!t^{-2})N$ bits in Regions D and E, respectively. Similarly, when the visible fraction dominates $v\!>\!1\!-\!t^{-2}$, we have $I(\secret;\visible)\!=\!S(\secret)$ (Regions C and E). In addition, the secret fully utilizes $M$ spins, with $S(\secret)\!=\!M$ bits, when $m\!<\!1\!-\!t^{-2}$ (Regions~A,\,B\,\&\,C), see Eq~\eqref{eq:marginal_entropy}.

That is, Regions A, B and C exhibit secret encoding at full capacity while allowing no, partial and perfect decoding, respectively~\footnote{Note that secret decoding can be computationally difficult even if it is information-theoretically allowed. Here, we focus on the information-theoretic properties of secret codes rather than the existence of efficient algorithms.}. This property is essential to information-theoretically secure secret sharing. In Region A, visible spins are too few, and zero information leakage guarantees absolute secrecy. In Region C, visible spins are adequately many, and vanishing \emph{a posteriori} entropy, $S(\secret\,|\,\visible)\!=\!S(\secret)-I(\secret;\visible)\!=\!0$, allows perfect inference of secret. However, this picture breaks down for long secrets $m\!>\!1\!-\!t^{-2}$ (Regions~D\,\&\,E) for which the information is always finite and secrecy is at best partial.

\begin{figure}
\centering
\includegraphics[width=\linewidth]{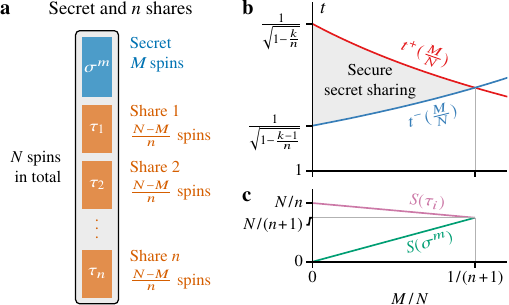}
\caption{\label{fig:kn}%
Phase diagram for threshold secret-sharing schemes, based on the random-energy model.
(a)~We split $N$ spins into an $M$-spin secret $\secret$ and $n$ shares $(\share_1,\share_2,\dots,\share_n)$, each with $(N\!-\!M)/n$ spins. 
(b)~The secrecy and correctness requirements of a $(k,n)$ threshold scheme result in temperature lower and upper bounds, $t^-$ and $t^+$ respectively, see Eqs~(\ref{eq:secrecy}-\ref{eq:t+}). These bounds depend on the secret length $m\!=\!M/N$ (as a proportion of the system size) and define a secure region (shaded area). Longer secrets are secure over a smaller temperature range. This range disappears completely ($t^-\!>\!t^+$) when $m\!>\!1/(n\!+\!1)$; that is, no secure scheme exists when the secret is longer than each share. 
(c) In the secure region, the entropy of the secret and of each share is equal to their lengths, i.e., $S(\secret)\!=\!M$ and $S(\share_i)\!=\!(N\!-\!M)/n$. We see that $S(\share_i)\!\ge\!S(\secret)$ with equality for $m\!=\!1/(n\!+\!1)$ at which $t^-\!=\!t^+\!=\!1/\sqrt{1\!-\!k/(n\!+\!1)}$.%
}
\end{figure}

These results form a basis for threshold secret-sharing schemes. So far, we consider the mutual information between the secret and visible spins where the visible fraction is taken to be continuous. We now turn to the case where the non-secret spins are divided into a finite number of shares and as a result, the visible fraction becomes discrete, taking only values that are multiples of the fraction of each share. As before, we reserve $M$ spins to encode the secret, $\secret$. But now we split the remaining $N\!-\!M$ spins, $\shares$, into $n$ equal shares---i.e., we decompose the system as follows, $\sigma\!=\!(\secret,\shares)\!=\!(\secret,\share_1,\share_2,\dots,\share_n)$ where $\share_i$ denotes the spin configuration of the \textit{i}\textsuperscript{\,th} share, see Fig~\ref{fig:kn}a. Here we also define $\share^{r}$ as any set of $r$ distinct shares.

A secure $(k,n)$ threshold scheme must meet two requirements (see, e.g., Ref~\cite{beimel:11}). First, \emph{secrecy} demands that fewer than $k$ shares leak no secret information,
\begin{equation}\label{eq:secrecy}
    I(\secret;\share^{k-1})=0.
\end{equation}
where $\share^{k-1}$ denotes $k\!-\!1$ shares \footnote{We need only consider the condition for $k\!-\!1$ shares since the data processing inequality implies that fewer shares cannot increase information. If $I(\secret;\share^{r})\!=\!0$ and $r\!>\!0$, then $I(\secret;\share^{r-1})\!=\!0$ due to nonnegativity of mutual information.}. Since each share amounts to a fraction $(1\!-\!m)/n$ of the whole system, access to $k\!-\!1$ shares is equivalent to visible spins $\visible$ with $v\!=\!(k-1)(1-m)/n$. From Eq~\eqref{eq:info} and Fig~\ref{fig:ternary}a-b, we see that the secrecy condition implies either
\begin{equation}\label{eq:t-}
    t\ge t^{-}(m)\equiv\frac{1}{\sqrt{(1-m)(1-(k-1)/n)}}, 
\end{equation}
or $t\!\le\!1$. But the latter corresponds to the frozen phase in which a secret carries no information, see Eq~\eqref{eq:marginal_entropy}. Second, \emph{correctness} requires that $k$ or more shares completely reveal the secret, 
\begin{equation}\label{eq:correctness}
    I(\secret;\share^{k})=S(\secret),
\end{equation}
or equivalently $S(\secret\,|\,\share^{k})\!=\!0$ \footnote{The condition for $k$ shares suffices since the data processing inequality implies that more shares cannot decrease information. If $I(\secret;\share^{r})\!=\!S(\secret)$ and $r\!<\!n$, then $I(\secret;\share^{r+1})\!=\!S(\secret)$ as the secret entropy bounds the information from above.}. Here, $\share^{k}$ is the same as $\visible$ with $v\!=\!k(1-m)/n$. Recalling Eq~\eqref{eq:info} (see Fig~\ref{fig:ternary}a-b), the above condition gives
\begin{equation}\label{eq:t+}
    t\le t^{+}(m)\equiv\frac{1}{\sqrt{1-k(1-m)/n}}.
\end{equation}
In Fig~\ref{fig:kn}b, we show that the secrecy and correctness requirements, Eqs~\eqref{eq:t-}\,\&\,\eqref{eq:t+}, define a secure region as we vary the temperature $t$ and secret length $m$. We see that this region is finite for $n\!\ge\!k\!>\!1$; we can always find a set of model parameters such that the REM implements a secure $(k,n)$ threshold scheme.

But the secret cannot be too long. In Fig~\ref{fig:kn}b, we also see that the temperature upper and lower bounds, $t^+(m)$ and $t^-(m)$, cross at $m\!=\!1/(n\!+\!1)$, thus ruling out secure schemes with a secret longer than this value. Indeed, this point is where the information contents of the secret and that of each share are equal. In Fig~\ref{fig:kn}c, we depict the entropy of the secret and that of each share in the secure region. We see that for $m\!<\!1/(n\!+\!1)$, each share has greater entropy than the secret, meaning that the secret amounts to only a fraction of the information in the share. At $m\!=\!1/(n\!+\!1)$ however, the secret and share entropies are equal and the secret encoding makes use of \emph{all} available information content in each share. Furthermore, in the secure regime, the entropies are at the physical limit, $M$ bits for the secret and $(N\!-\!M)/n$ bits for each share. In this sense, REM secret sharing at $m\!=\!1/(n\!+\!1)$ is optimal. This optimality also requires a specific temperature (see Fig~\ref{fig:kn}c),
\begin{equation}
t^*=\frac{1}{\sqrt{1-k/(n+1)}},
\end{equation}
which, quite remarkably, exists for any $(k,n)$ threshold scheme with $1\!\le\!k\!\le\!n$.

\begin{figure}
\centering
\includegraphics[width=\linewidth]{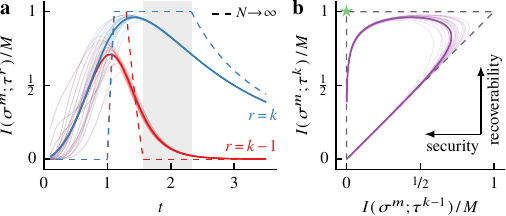}
\caption{\label{fig:info_vs_t}%
Finite-spin REMs implement secret-sharing schemes with a tradeoff between security and recoverability. 
(a)~We show the mutual information between the secret $\secret$ and $r$ distinct shares $\share^r$ for the cases where $r$ is at the threshold (blue) and subthreshold (red). The thick lines are the average of 20 independent realizations of the REM, each depicted by thin lines. The dashed lines correspond to the thermodynamic limit ($N\!\to\!\infty$) and the shaded area to the temperature range which satisfies the requirements of a secure scheme, see Eqs~(\ref{eq:secrecy}-\ref{eq:t+}). 
(b)~We display the parametric curve of the at-threshold and subthreshold information terms, (vertical and horizontal axes, respectively). The thick curve is the average of 20 independent REMs (thin curves), and the dashed line is the corresponding parametric plot for the thermodynamic limit. This plot illustrates how far finite-spin REM secret sharing is from an information-theoretically secure scheme~($\star$). 
Here $N\!=\!27$, $M\!=\!5$ and $(k,n)\!=\!(2,2)$. For the thermodynamic case, we let $N\!\to\!\infty$ while fixing the ratio $M/N$.%
}
\end{figure}

So far we have shown that the REM in the thermodynamic limit ($N\!\to\!\infty$) can implement a threshold secret-sharing scheme that is both secure and optimal. We now demonstrate that a finite-spin REM displays similar behaviors, albeit with relaxed secrecy and correctness guarantees.

In Fig~\ref{fig:info_vs_t}a, we depict the temperature dependence of the mutual information between the secret $\secret$ and $r$ shares $\share^r$ when the number of shares is at and below the threshold ($r\!=\!k$ and $r\!=\!k\!-\!1$, respectively). The dashed lines show the result for the thermodynamic case, which corresponds to sending $N\!\to\!\infty$ while fixing $m\!=\!M/N$. In this case, the information vanishes in the frozen phase, increases as the temperature rises above the critical point, plateaus at intermediate temperatures, and decays until it vanishes again at an adequately high temperature, see Eq~\eqref{eq:info}. The temperature range over which the information plateaus and the point at which it vanishes depend on the number of available shares $r$. This behavior yields a range of temperatures (shaded region) in which the REM meets both secrecy and correctness requirements, Eqs~\eqref{eq:secrecy}\,\&\,\eqref{eq:correctness}.

For the finite-system case, we consider $N\!=\!27$ and $M\!=\!5$, and depict the average information (thick lines) for 20 REMs (thin lines), each with independently generated energy levels. We see that the information exhibits temperature dependence that is qualitatively similar to the thermodynamic case, but the sharp features become smooth. In particular, a finite system does not completely freeze at low temperatures and the information remains finite below the thermodynamic critical temperature. Still, at intermediate temperatures (shaded region), a significant information difference exists between the cases where the number of shares is at and below the threshold ($r\!=\!k$ and $r\!=\!k\!-\!1$, respectively). In other words, enough shares ($r\!\ge\!k$) reveal most of the secret, but subthreshold shares ($r\!<\!k$) tell us very little about the secret. This behavior captures the essence of a secret-sharing scheme, albeit without strict secrecy and correctness guarantees. Indeed, a finite-spin REM exhibits a tradeoff between secrecy and recoverability.

In Fig~\ref{fig:info_vs_t}b, we illustrate this tradeoff via the parametric curve of the information between the secret and $k$ shares (vertical axis) and between the secret and $k\!-\!1$ shares (horizontal axis). The former, $I(\secret;\share^{k})$, measures the recoverability of the secret given access to enough shares whereas the latter, $I(\secret;\share^{k-1})$, quantifies the deviation from perfect secrecy when accessible shares fall below the threshold. Both information terms vanish at zero temperature. Raising the temperature from $t\!=\!0$ increases both information at approximately the same rate, $I(\secret;\share^{k})\!\approx\!I(\secret;\share^{k-1})$ at small $t$, suggesting a redundant coding regime. That is, each share encodes the same bits about the secret, therefore an additional share provides little extra information about the secret. But $I(\secret;\share^{k-1})$ reaches its maximum and begins to decay with $t$ while $I(\secret;\share^{k})$ still grows, resulting in enhanced secrecy and recoverability. This mutual enhancement signifies the onset of synergistic coding; however, it only occurs for a limited temperature range (between the maxima of $I(\secret;\share^{k})$ and $I(\secret;\share^{k-1})$, see Fig~\ref{fig:info_vs_t}a). As the temperature rises further, $I(\secret;\share^{k})$, and recoverability, eventually peaks and decreases while $I(\secret;\share^{k-1})$ approaches zero. In this regime, an increase in security is accompanied by a decrease in recoverability and vice versa. Importantly, this tradeoff defines the optimal frontier of maximum recoverability for each value of security budgets.

In the thermodynamic limit, this frontier collapses into a single point that satisfies the requirements of a secure secret-sharing scheme, $I(\secret;\share^{k})\!=\!M$ and $I(\secret;\share^{k-1})\!=\!0$ [see Eqs~\eqref{eq:correctness}\,\&\,\eqref{eq:secrecy}]. We note that this point corresponds to not one but a range of temperatures (i.e., shaded range in Fig~\ref{fig:info_vs_t}a and shaded region in Fig~\ref{fig:kn}).

\begin{figure}
\centering
\includegraphics[width=\linewidth]{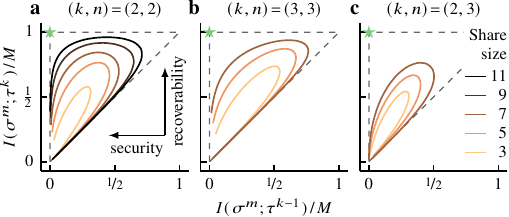}
\caption{\label{fig:info_plane}%
Larger REMs yield more secure secret-sharing schemes. We show the parametric curves of the at-threshold and subthreshold information terms (vertical and horizontal axes, respectively) for a five-spin secret ($M\!=\!5$) under three threshold schemes, $(k,n)\!=\!(2,2),(3,3),(2,3)$ (left to right) at various share size (see legend). In all cases, the tradeoff frontiers become closer to the ideal scheme ($\star$) with increasing share size. Here the information is the average of 20 independent REMs.%
}
\end{figure}

In finite systems, improving the optimal frontier of random-energy secret sharing requires increasing the system size. In Fig~\ref{fig:info_plane}, we consider a fixed secret size of $M\!=\!5$ and compare the optimal frontiers (averaged over 20 independent REMs) for different numbers of spins in each share (see legend) for threshold schemes with $(k,n)\!=\!(2,2),(3,3),(2,3)$ (left to right). In all cases, we see that as share size grows and the system becomes larger, the optimal frontier edges closer to the ideal secret-sharing scheme.

To summarize, we characterize the thermodynamics of information in the REM and show that the synergistic correlations among spins provide a basis for a secret-sharing scheme. We further demonstrate that this scheme is information-theoretically secure and optimal. Our analytical results for the thermodynamic limit are in good qualitative agreement with the numerical simulations of finite-spin REMs. Our work represents a further example of the curious statistical properties of the REM and highlights information theory as a unifying element of problems in physics and computer science. The mapping between secret sharing and the exactly solvable REM may facilitate the application of statistical physics techniques in cryptography and inspire novel physical models based on secret-sharing schemes.

\begin{acknowledgments}
This work was supported in part by the National Science Foundation, through the Center for the Physics of Biological Function (PHY-1734030), the Simons Foundation and the Sloan Foundation.
\end{acknowledgments}


\begin{thebibliography}{32}%
\makeatletter
\providecommand \@ifxundefined [1]{%
 \@ifx{#1\undefined}
}%
\providecommand \@ifnum [1]{%
 \ifnum #1\expandafter \@firstoftwo
 \else \expandafter \@secondoftwo
 \fi
}%
\providecommand \@ifx [1]{%
 \ifx #1\expandafter \@firstoftwo
 \else \expandafter \@secondoftwo
 \fi
}%
\providecommand \natexlab [1]{#1}%
\providecommand \enquote  [1]{``#1''}%
\providecommand \bibnamefont  [1]{#1}%
\providecommand \bibfnamefont [1]{#1}%
\providecommand \citenamefont [1]{#1}%
\providecommand \href@noop [0]{\@secondoftwo}%
\providecommand \href [0]{\begingroup \@sanitize@url \@href}%
\providecommand \@href[1]{\@@startlink{#1}\@@href}%
\providecommand \@@href[1]{\endgroup#1\@@endlink}%
\providecommand \@sanitize@url [0]{\catcode `\\12\catcode `\$12\catcode
  `\&12\catcode `\#12\catcode `\^12\catcode `\_12\catcode `\%12\relax}%
\providecommand \@@startlink[1]{}%
\providecommand \@@endlink[0]{}%
\providecommand \url  [0]{\begingroup\@sanitize@url \@url }%
\providecommand \@url [1]{\endgroup\@href {#1}{\urlprefix }}%
\providecommand \urlprefix  [0]{URL }%
\providecommand \Eprint [0]{\href }%
\providecommand \doibase [0]{https://doi.org/}%
\providecommand \selectlanguage [0]{\@gobble}%
\providecommand \bibinfo  [0]{\@secondoftwo}%
\providecommand \bibfield  [0]{\@secondoftwo}%
\providecommand \translation [1]{[#1]}%
\providecommand \BibitemOpen [0]{}%
\providecommand \bibitemStop [0]{}%
\providecommand \bibitemNoStop [0]{.\EOS\space}%
\providecommand \EOS [0]{\spacefactor3000\relax}%
\providecommand \BibitemShut  [1]{\csname bibitem#1\endcsname}%
\let\auto@bib@innerbib\@empty
\bibitem [{\citenamefont {Shamir}(1979)}]{shamir:79}%
  \BibitemOpen
  \bibfield  {author} {\bibinfo {author} {\bibfnamefont {A.}~\bibnamefont
  {Shamir}},\ }\bibfield  {title} {\bibinfo {title} {How to share a secret},\
  }\href {https://doi.org/10.1145/359168.359176} {\bibfield  {journal}
  {\bibinfo  {journal} {Commun.\ ACM}\ }\textbf {\bibinfo {volume} {22}},\
  \bibinfo {pages} {612} (\bibinfo {year} {1979})}\BibitemShut {NoStop}%
\bibitem [{\citenamefont {Blakley}(1979)}]{blakley:79}%
  \BibitemOpen
  \bibfield  {author} {\bibinfo {author} {\bibfnamefont {G.~R.}\ \bibnamefont
  {Blakley}},\ }\bibfield  {title} {\bibinfo {title} {Safeguarding
  cryptographic keys},\ }in\ \href {https://doi.org/10.1109/MARK.1979.8817296}
  {\emph {\bibinfo {booktitle} {1979 International Workshop on Managing
  Requirements Knowledge (MARK)}}}\ (\bibinfo {year} {1979})\ pp.\ \bibinfo
  {pages} {313--318}\BibitemShut {NoStop}%
\bibitem [{\citenamefont {Beimel}(2011)}]{beimel:11}%
  \BibitemOpen
  \bibfield  {author} {\bibinfo {author} {\bibfnamefont {A.}~\bibnamefont
  {Beimel}},\ }\bibfield  {title} {\bibinfo {title} {Secret-sharing schemes: A
  survey},\ }in\ \href {https://doi.org/10.1007/978-3-642-20901-7_2} {\emph
  {\bibinfo {booktitle} {Coding and Cryptology}}},\ \bibinfo {editor} {edited
  by\ \bibinfo {editor} {\bibfnamefont {Y.~M.}\ \bibnamefont {Chee}}, \bibinfo
  {editor} {\bibfnamefont {Z.}~\bibnamefont {Guo}}, \bibinfo {editor}
  {\bibfnamefont {S.}~\bibnamefont {Ling}}, \bibinfo {editor} {\bibfnamefont
  {F.}~\bibnamefont {Shao}}, \bibinfo {editor} {\bibfnamefont {Y.}~\bibnamefont
  {Tang}}, \bibinfo {editor} {\bibfnamefont {H.}~\bibnamefont {Wang}},\ and\
  \bibinfo {editor} {\bibfnamefont {C.}~\bibnamefont {Xing}}}\ (\bibinfo
  {publisher} {Springer},\ \bibinfo {address} {Berlin, Heidelberg},\ \bibinfo
  {year} {2011})\ pp.\ \bibinfo {pages} {11--46}\BibitemShut {NoStop}%
\bibitem [{\citenamefont {Balasubramanian}\ \emph {et~al.}(2021)\citenamefont
  {Balasubramanian}, \citenamefont {Kar}, \citenamefont {Parrikar},
  \citenamefont {S{\'a}rosi},\ and\ \citenamefont
  {Ugajin}}]{balasubramanian:21}%
  \BibitemOpen
  \bibfield  {author} {\bibinfo {author} {\bibfnamefont {V.}~\bibnamefont
  {Balasubramanian}}, \bibinfo {author} {\bibfnamefont {A.}~\bibnamefont
  {Kar}}, \bibinfo {author} {\bibfnamefont {O.}~\bibnamefont {Parrikar}},
  \bibinfo {author} {\bibfnamefont {G.}~\bibnamefont {S{\'a}rosi}},\ and\
  \bibinfo {author} {\bibfnamefont {T.}~\bibnamefont {Ugajin}},\ }\bibfield
  {title} {\bibinfo {title} {Geometric secret sharing in a model of hawking
  radiation},\ }\href {https://doi.org/10.1007/JHEP01(2021)177} {\bibfield
  {journal} {\bibinfo  {journal} {J. High Energy Phys.}\ }\textbf {\bibinfo
  {volume} {2021}}\bibinfo  {number} { (1)},\ \bibinfo {pages}
  {177}}\BibitemShut {NoStop}%
\bibitem [{\citenamefont {Brenner}\ \emph {et~al.}(2000)\citenamefont
  {Brenner}, \citenamefont {Strong}, \citenamefont {Koberle}, \citenamefont
  {Bialek},\ and\ \citenamefont {de~Ruyter~van Steveninck}}]{brenner:00}%
  \BibitemOpen
\bibfield  {number} {  }\bibfield  {author} {\bibinfo {author} {\bibfnamefont
  {N.}~\bibnamefont {Brenner}}, \bibinfo {author} {\bibfnamefont {S.~P.}\
  \bibnamefont {Strong}}, \bibinfo {author} {\bibfnamefont {R.}~\bibnamefont
  {Koberle}}, \bibinfo {author} {\bibfnamefont {W.}~\bibnamefont {Bialek}},\
  and\ \bibinfo {author} {\bibfnamefont {R.~R.}\ \bibnamefont {de~Ruyter~van
  Steveninck}},\ }\bibfield  {title} {\bibinfo {title} {Synergy in a neural
  code},\ }\href {https://doi.org/10.1162/089976600300015259} {\bibfield
  {journal} {\bibinfo  {journal} {Neural Comput.}\ }\textbf {\bibinfo {volume}
  {12}},\ \bibinfo {pages} {1531} (\bibinfo {year} {2000})}\BibitemShut
  {NoStop}%
\bibitem [{\citenamefont {Schneidman}\ \emph {et~al.}(2003)\citenamefont
  {Schneidman}, \citenamefont {Bialek},\ and\ \citenamefont
  {Berry~II}}]{schneidman:03b}%
  \BibitemOpen
  \bibfield  {author} {\bibinfo {author} {\bibfnamefont {E.}~\bibnamefont
  {Schneidman}}, \bibinfo {author} {\bibfnamefont {W.}~\bibnamefont {Bialek}},\
  and\ \bibinfo {author} {\bibfnamefont {M.~J.}\ \bibnamefont {Berry~II}},\
  }\bibfield  {title} {\bibinfo {title} {Synergy, redundancy, and independence
  in population codes},\ }\href
  {https://doi.org/10.1523/JNEUROSCI.23-37-11539.2003} {\bibfield  {journal}
  {\bibinfo  {journal} {J. Neurosci.}\ }\textbf {\bibinfo {volume} {23}},\
  \bibinfo {pages} {11539} (\bibinfo {year} {2003})}\BibitemShut {NoStop}%
\bibitem [{\citenamefont {Derrida}(1980)}]{derrida:80}%
  \BibitemOpen
  \bibfield  {author} {\bibinfo {author} {\bibfnamefont {B.}~\bibnamefont
  {Derrida}},\ }\bibfield  {title} {\bibinfo {title} {Random-energy model:
  Limit of a family of disordered models},\ }\href
  {https://doi.org/10.1103/PhysRevLett.45.79} {\bibfield  {journal} {\bibinfo
  {journal} {Phys. Rev. Lett.}\ }\textbf {\bibinfo {volume} {45}},\ \bibinfo
  {pages} {79} (\bibinfo {year} {1980})}\BibitemShut {NoStop}%
\bibitem [{\citenamefont {Derrida}(1981)}]{derrida:81}%
  \BibitemOpen
  \bibfield  {author} {\bibinfo {author} {\bibfnamefont {B.}~\bibnamefont
  {Derrida}},\ }\bibfield  {title} {\bibinfo {title} {Random-energy model: An
  exactly solvable model of disordered systems},\ }\href
  {https://doi.org/10.1103/PhysRevB.24.2613} {\bibfield  {journal} {\bibinfo
  {journal} {Phys. Rev. B}\ }\textbf {\bibinfo {volume} {24}},\ \bibinfo
  {pages} {2613} (\bibinfo {year} {1981})}\BibitemShut {NoStop}%
\bibitem [{\citenamefont {Bryngelson}\ and\ \citenamefont
  {Wolynes}(1987)}]{bryngelson:87}%
  \BibitemOpen
  \bibfield  {author} {\bibinfo {author} {\bibfnamefont {J.~D.}\ \bibnamefont
  {Bryngelson}}\ and\ \bibinfo {author} {\bibfnamefont {P.~G.}\ \bibnamefont
  {Wolynes}},\ }\bibfield  {title} {\bibinfo {title} {Spin glasses and the
  statistical mechanics of protein folding.},\ }\href
  {https://doi.org/10.1073/pnas.84.21.7524} {\bibfield  {journal} {\bibinfo
  {journal} {Proc. Natl. Acad. Sci. USA}\ }\textbf {\bibinfo {volume} {84}},\
  \bibinfo {pages} {7524} (\bibinfo {year} {1987})}\BibitemShut {NoStop}%
\bibitem [{\citenamefont {Shakhnovich}\ and\ \citenamefont
  {Gutin}(1990)}]{shakhnovich:90}%
  \BibitemOpen
  \bibfield  {author} {\bibinfo {author} {\bibfnamefont {E.~I.}\ \bibnamefont
  {Shakhnovich}}\ and\ \bibinfo {author} {\bibfnamefont {A.~M.}\ \bibnamefont
  {Gutin}},\ }\bibfield  {title} {\bibinfo {title} {Implications of
  thermodynamics of protein folding for evolution of primary sequences},\
  }\href {https://doi.org/10.1038/346773a0} {\bibfield  {journal} {\bibinfo
  {journal} {Nature}\ }\textbf {\bibinfo {volume} {346}},\ \bibinfo {pages}
  {773} (\bibinfo {year} {1990})}\BibitemShut {NoStop}%
\bibitem [{\citenamefont {Pande}\ \emph {et~al.}(1996)\citenamefont {Pande},
  \citenamefont {Grosberg}, \citenamefont {Joerg},\ and\ \citenamefont
  {Tanaka}}]{pande:96}%
  \BibitemOpen
  \bibfield  {author} {\bibinfo {author} {\bibfnamefont {V.~S.}\ \bibnamefont
  {Pande}}, \bibinfo {author} {\bibfnamefont {A.~Y.}\ \bibnamefont {Grosberg}},
  \bibinfo {author} {\bibfnamefont {C.}~\bibnamefont {Joerg}},\ and\ \bibinfo
  {author} {\bibfnamefont {T.}~\bibnamefont {Tanaka}},\ }\bibfield  {title}
  {\bibinfo {title} {Is heteropolymer freezing well described by the random
  energy model?},\ }\href {https://doi.org/10.1103/PhysRevLett.76.3987}
  {\bibfield  {journal} {\bibinfo  {journal} {Phys. Rev. Lett.}\ }\textbf
  {\bibinfo {volume} {76}},\ \bibinfo {pages} {3987} (\bibinfo {year}
  {1996})}\BibitemShut {NoStop}%
\bibitem [{\citenamefont {Hormoz}\ and\ \citenamefont
  {Brenner}(2011)}]{hormoz:11}%
  \BibitemOpen
  \bibfield  {author} {\bibinfo {author} {\bibfnamefont {S.}~\bibnamefont
  {Hormoz}}\ and\ \bibinfo {author} {\bibfnamefont {M.~P.}\ \bibnamefont
  {Brenner}},\ }\bibfield  {title} {\bibinfo {title} {Design principles for
  self-assembly with short-range interactions},\ }\href
  {https://doi.org/10.1073/pnas.1014094108} {\bibfield  {journal} {\bibinfo
  {journal} {Proc. Natl. Acad. Sci. USA}\ }\textbf {\bibinfo {volume} {108}},\
  \bibinfo {pages} {5193} (\bibinfo {year} {2011})}\BibitemShut {NoStop}%
\bibitem [{\citenamefont {Fisher}\ and\ \citenamefont
  {Mehta}(2014)}]{fisher:14}%
  \BibitemOpen
  \bibfield  {author} {\bibinfo {author} {\bibfnamefont {C.~K.}\ \bibnamefont
  {Fisher}}\ and\ \bibinfo {author} {\bibfnamefont {P.}~\bibnamefont {Mehta}},\
  }\bibfield  {title} {\bibinfo {title} {The transition between the niche and
  neutral regimes in ecology},\ }\href
  {https://doi.org/10.1073/pnas.1405637111} {\bibfield  {journal} {\bibinfo
  {journal} {Proc. Natl. Acad. Sci. USA}\ }\textbf {\bibinfo {volume} {111}},\
  \bibinfo {pages} {13111} (\bibinfo {year} {2014})}\BibitemShut {NoStop}%
\bibitem [{\citenamefont {Laumann}\ \emph {et~al.}(2014)\citenamefont
  {Laumann}, \citenamefont {Pal},\ and\ \citenamefont
  {Scardicchio}}]{laumann:14}%
  \BibitemOpen
  \bibfield  {author} {\bibinfo {author} {\bibfnamefont {C.~R.}\ \bibnamefont
  {Laumann}}, \bibinfo {author} {\bibfnamefont {A.}~\bibnamefont {Pal}},\ and\
  \bibinfo {author} {\bibfnamefont {A.}~\bibnamefont {Scardicchio}},\
  }\bibfield  {title} {\bibinfo {title} {Many-body mobility edge in a
  mean-field quantum spin glass},\ }\href
  {https://doi.org/10.1103/PhysRevLett.113.200405} {\bibfield  {journal}
  {\bibinfo  {journal} {Phys. Rev. Lett.}\ }\textbf {\bibinfo {volume} {113}},\
  \bibinfo {pages} {200405} (\bibinfo {year} {2014})}\BibitemShut {NoStop}%
\bibitem [{\citenamefont {Lazarides}\ \emph {et~al.}(2015)\citenamefont
  {Lazarides}, \citenamefont {Das},\ and\ \citenamefont
  {Moessner}}]{lazarides:15}%
  \BibitemOpen
  \bibfield  {author} {\bibinfo {author} {\bibfnamefont {A.}~\bibnamefont
  {Lazarides}}, \bibinfo {author} {\bibfnamefont {A.}~\bibnamefont {Das}},\
  and\ \bibinfo {author} {\bibfnamefont {R.}~\bibnamefont {Moessner}},\
  }\bibfield  {title} {\bibinfo {title} {Fate of many-body localization under
  periodic driving},\ }\href {https://doi.org/10.1103/PhysRevLett.115.030402}
  {\bibfield  {journal} {\bibinfo  {journal} {Phys. Rev. Lett.}\ }\textbf
  {\bibinfo {volume} {115}},\ \bibinfo {pages} {030402} (\bibinfo {year}
  {2015})}\BibitemShut {NoStop}%
\bibitem [{\citenamefont {Baldwin}\ \emph {et~al.}(2016)\citenamefont
  {Baldwin}, \citenamefont {Laumann}, \citenamefont {Pal},\ and\ \citenamefont
  {Scardicchio}}]{baldwin:16}%
  \BibitemOpen
  \bibfield  {author} {\bibinfo {author} {\bibfnamefont {C.~L.}\ \bibnamefont
  {Baldwin}}, \bibinfo {author} {\bibfnamefont {C.~R.}\ \bibnamefont
  {Laumann}}, \bibinfo {author} {\bibfnamefont {A.}~\bibnamefont {Pal}},\ and\
  \bibinfo {author} {\bibfnamefont {A.}~\bibnamefont {Scardicchio}},\
  }\bibfield  {title} {\bibinfo {title} {The many-body localized phase of the
  quantum random energy model},\ }\href
  {https://doi.org/10.1103/PhysRevB.93.024202} {\bibfield  {journal} {\bibinfo
  {journal} {Phys. Rev. B}\ }\textbf {\bibinfo {volume} {93}},\ \bibinfo
  {pages} {024202} (\bibinfo {year} {2016})}\BibitemShut {NoStop}%
\bibitem [{\citenamefont {Sourlas}(1989)}]{sourlas:89}%
  \BibitemOpen
  \bibfield  {author} {\bibinfo {author} {\bibfnamefont {N.}~\bibnamefont
  {Sourlas}},\ }\bibfield  {title} {\bibinfo {title} {Spin-glass models as
  error-correcting codes},\ }\href {https://doi.org/10.1038/339693a0}
  {\bibfield  {journal} {\bibinfo  {journal} {Nature}\ }\textbf {\bibinfo
  {volume} {339}},\ \bibinfo {pages} {693} (\bibinfo {year}
  {1989})}\BibitemShut {NoStop}%
\bibitem [{\citenamefont {Mertens}(2000)}]{mertens:00}%
  \BibitemOpen
  \bibfield  {author} {\bibinfo {author} {\bibfnamefont {S.}~\bibnamefont
  {Mertens}},\ }\bibfield  {title} {\bibinfo {title} {Random costs in
  combinatorial optimization},\ }\href
  {https://doi.org/10.1103/PhysRevLett.84.1347} {\bibfield  {journal} {\bibinfo
   {journal} {Phys. Rev. Lett.}\ }\textbf {\bibinfo {volume} {84}},\ \bibinfo
  {pages} {1347} (\bibinfo {year} {2000})}\BibitemShut {NoStop}%
\bibitem [{\citenamefont {Borgs}\ \emph {et~al.}(2001)\citenamefont {Borgs},
  \citenamefont {Chayes},\ and\ \citenamefont {Pittel}}]{borgs:01}%
  \BibitemOpen
  \bibfield  {author} {\bibinfo {author} {\bibfnamefont {C.}~\bibnamefont
  {Borgs}}, \bibinfo {author} {\bibfnamefont {J.}~\bibnamefont {Chayes}},\ and\
  \bibinfo {author} {\bibfnamefont {B.}~\bibnamefont {Pittel}},\ }\bibfield
  {title} {\bibinfo {title} {Phase transition and finite-size scaling for the
  integer partitioning problem},\ }\href {https://doi.org/10.1002/rsa.10004}
  {\bibfield  {journal} {\bibinfo  {journal} {Random Struct. Algorithms}\
  }\textbf {\bibinfo {volume} {19}},\ \bibinfo {pages} {247} (\bibinfo {year}
  {2001})}\BibitemShut {NoStop}%
\bibitem [{\citenamefont {M\'ezard}\ and\ \citenamefont
  {Montanari}(2009)}]{mezard:09}%
  \BibitemOpen
  \bibfield  {author} {\bibinfo {author} {\bibfnamefont {M.}~\bibnamefont
  {M\'ezard}}\ and\ \bibinfo {author} {\bibfnamefont {A.}~\bibnamefont
  {Montanari}},\ }\href
  {https://doi.org/10.1093/acprof:oso/9780198570837.001.0001} {\emph {\bibinfo
  {title} {Information, Physics, and Computation}}}\ (\bibinfo  {publisher}
  {Oxford University Press},\ \bibinfo {year} {2009})\BibitemShut {NoStop}%
\bibitem [{Note1()}]{Note1}%
  \BibitemOpen
  \bibinfo {note} {We slightly abuse the notation and use $\sigma $ to denote
  both random variables and their realizations.}\BibitemShut {Stop}%
\bibitem [{Note2()}]{Note2}%
  \BibitemOpen
  \bibinfo {note} {\label {ft:marginal_entropy}The argument, leading to this
  expression, extends to any extensive subsystem of the REM; in particular,
  Eq~\protect \eqref {eq:marginal_entropy} gives the marginal entropy of a
  subsystem that is a fraction $m$ of the full system.}\BibitemShut {Stop}%
\bibitem [{Note3()}]{Note3}%
  \BibitemOpen
  \bibinfo {note} {The mutual information between two equal halves of the REM
  \cite {bialek:20} is a special case of our result, with $h\protect
  \!=\protect \!0$ and $m\protect \!=\protect \!v\protect \!=\protect
  \!1/2$.}\BibitemShut {Stop}%
\bibitem [{\citenamefont {Grassberger}(1986)}]{grassberger:86}%
  \BibitemOpen
  \bibfield  {author} {\bibinfo {author} {\bibfnamefont {P.}~\bibnamefont
  {Grassberger}},\ }\bibfield  {title} {\bibinfo {title} {Toward a quantitative
  theory of self-generated complexity},\ }\href
  {https://doi.org/10.1007/BF00668821} {\bibfield  {journal} {\bibinfo
  {journal} {Int. J. Theor. Phys.}\ }\textbf {\bibinfo {volume} {25}},\
  \bibinfo {pages} {907} (\bibinfo {year} {1986})}\BibitemShut {NoStop}%
\bibitem [{\citenamefont {Bialek}\ \emph {et~al.}(2001)\citenamefont {Bialek},
  \citenamefont {Nemenman},\ and\ \citenamefont {Tishby}}]{bialek:01}%
  \BibitemOpen
  \bibfield  {author} {\bibinfo {author} {\bibfnamefont {W.}~\bibnamefont
  {Bialek}}, \bibinfo {author} {\bibfnamefont {I.}~\bibnamefont {Nemenman}},\
  and\ \bibinfo {author} {\bibfnamefont {N.}~\bibnamefont {Tishby}},\
  }\bibfield  {title} {\bibinfo {title} {Predictability, complexity, and
  learning},\ }\href {https://doi.org/10.1162/089976601753195969} {\bibfield
  {journal} {\bibinfo  {journal} {Neural Comput.}\ }\textbf {\bibinfo {volume}
  {13}},\ \bibinfo {pages} {2409} (\bibinfo {year} {2001})}\BibitemShut
  {NoStop}%
\bibitem [{\citenamefont {Wilms}\ \emph {et~al.}(2012)\citenamefont {Wilms},
  \citenamefont {Vidal}, \citenamefont {Verstraete},\ and\ \citenamefont
  {Dusuel}}]{wilms:12}%
  \BibitemOpen
  \bibfield  {author} {\bibinfo {author} {\bibfnamefont {J.}~\bibnamefont
  {Wilms}}, \bibinfo {author} {\bibfnamefont {J.}~\bibnamefont {Vidal}},
  \bibinfo {author} {\bibfnamefont {F.}~\bibnamefont {Verstraete}},\ and\
  \bibinfo {author} {\bibfnamefont {S.}~\bibnamefont {Dusuel}},\ }\bibfield
  {title} {\bibinfo {title} {Finite-temperature mutual information in a simple
  phase transition},\ }\href {https://doi.org/10.1088/1742-5468/2012/01/P01023}
  {\bibfield  {journal} {\bibinfo  {journal} {J. Stat. Mech.}\ }\textbf
  {\bibinfo {volume} {2012}},\ \bibinfo {pages} {P01023} (\bibinfo {year}
  {2012})}\BibitemShut {NoStop}%
\bibitem [{\citenamefont {Cohen}\ \emph {et~al.}(2015)\citenamefont {Cohen},
  \citenamefont {Rittenberg},\ and\ \citenamefont {Sadhu}}]{cohen:15}%
  \BibitemOpen
  \bibfield  {author} {\bibinfo {author} {\bibfnamefont {O.}~\bibnamefont
  {Cohen}}, \bibinfo {author} {\bibfnamefont {V.}~\bibnamefont {Rittenberg}},\
  and\ \bibinfo {author} {\bibfnamefont {T.}~\bibnamefont {Sadhu}},\ }\bibfield
   {title} {\bibinfo {title} {Shared information in classical mean-field
  models},\ }\href {https://doi.org/10.1088/1751-8113/48/5/055002} {\bibfield
  {journal} {\bibinfo  {journal} {J. Phys. A}\ }\textbf {\bibinfo {volume}
  {48}},\ \bibinfo {pages} {055002} (\bibinfo {year} {2015})}\BibitemShut
  {NoStop}%
\bibitem [{\citenamefont {Ngampruetikorn}\ \emph {et~al.}(2023)\citenamefont
  {Ngampruetikorn}, \citenamefont {Nemenman},\ and\ \citenamefont
  {Schwab}}]{ngampruetikorn:23}%
  \BibitemOpen
  \bibfield  {author} {\bibinfo {author} {\bibfnamefont {V.}~\bibnamefont
  {Ngampruetikorn}}, \bibinfo {author} {\bibfnamefont {I.}~\bibnamefont
  {Nemenman}},\ and\ \bibinfo {author} {\bibfnamefont {D.~J.}\ \bibnamefont
  {Schwab}},\ }\bibfield  {title} {\bibinfo {title} {{Extrinsic vs Intrinsic
  Criticality in Systems with Many Components}}} (\bibinfo {year} {2023}),\
  \bibinfo {note} {in preparation}\BibitemShut {NoStop}%
\bibitem [{\citenamefont {Bialek}\ \emph {et~al.}(2020)\citenamefont {Bialek},
  \citenamefont {Palmer},\ and\ \citenamefont {Schwab}}]{bialek:20}%
  \BibitemOpen
  \bibfield  {author} {\bibinfo {author} {\bibfnamefont {W.}~\bibnamefont
  {Bialek}}, \bibinfo {author} {\bibfnamefont {S.~E.}\ \bibnamefont {Palmer}},\
  and\ \bibinfo {author} {\bibfnamefont {D.~J.}\ \bibnamefont {Schwab}},\
  }\href@noop {} {\bibinfo {title} {What makes it possible to learn probability
  distributions in the natural world?}} (\bibinfo {year} {2020}),\ \Eprint
  {https://arxiv.org/abs/2008.12279} {arXiv:2008.12279 [cond-mat.stat-mech]}
  \BibitemShut {NoStop}%
\bibitem [{Note4()}]{Note4}%
  \BibitemOpen
  \bibinfo {note} {Note that secret decoding can be computationally difficult
  even if it is information-theoretically allowed. Here, we focus on the
  information-theoretic properties of secret codes rather than the existence of
  efficient algorithms.}\BibitemShut {Stop}%
\bibitem [{Note5()}]{Note5}%
  \BibitemOpen
  \bibinfo {note} {We need only consider the condition for $k\protect
  \!-\protect \!1$ shares since the data processing inequality implies that
  fewer shares cannot increase information. If $I(\sigma ^m;\tau ^{r})\protect
  \!=\protect \!0$ and $r\protect \!>\protect \!0$, then $I(\sigma ^m;\tau
  ^{r-1})\protect \!=\protect \!0$ due to nonnegativity of mutual
  information.}\BibitemShut {Stop}%
\bibitem [{Note6()}]{Note6}%
  \BibitemOpen
  \bibinfo {note} {The condition for $k$ shares suffices since the data
  processing inequality implies that more shares cannot decrease information.
  If $I(\sigma ^m;\tau ^{r})\protect \!=\protect \!S(\sigma ^m)$ and $r\protect
  \!<\protect \!n$, then $I(\sigma ^m;\tau ^{r+1})\protect \!=\protect
  \!S(\sigma ^m)$ as the secret entropy bounds the information from
  above.}\BibitemShut {Stop}%
\end{thebibliography}
%

\end{document}